\documentclass[vecphys]{svmult}

\usepackage{makeidx}         
\usepackage{graphicx}        
\usepackage{multicol}        
\usepackage[bottom]{footmisc}

\makeindex             

\usepackage{amssymb}

\begin{document}

\title*{Quasicrystals}
\author{Uwe Grimm\inst{1}\and
Peter Kramer\inst{2}}
\institute{Department of Mathematics, The Open University,
Walton Hall, Milton Keynes MK7 6AA, UK.
Email: \texttt{u.g.grimm@open.ac.uk}
\and Institut f\"{u}r Theoretische Physik, Universit\"{a}t T\"{u}bingen,
72076 T\"{u}bingen, Germany.
Email: \texttt{peter.kramer@uni-tuebingen.de}}
\maketitle

\begin{abstract}
Mathematicians have been interested in non-periodic tilings of space
for decades; however, it was the unexpected discovery of
non-periodically ordered structures in intermetallic alloys which
brought this subject into the limelight. These fascinating materials,
now called quasicrystals, are characterised by the coexistence of
long-range atomic order and `forbidden' symmetries which are
incompatible with periodic arrangements in three-dimensional space.
In the first part of this review, we summarise the main properties of
quasicrystals, and describe how their structures relate to
non-periodic tilings of space. The celebrated Penrose and
Ammann-Beenker tilings are introduced as illustrative examples. The
second part provides a closer look at the underlying mathematics. 
Starting from Bohr's theory of quasiperiodic functions, a
general framework for constructing non-periodic tilings of space is
described, and an alternative description as quasiperiodic coverings by
overlapping clusters is discussed.
\end{abstract}

\section{Aperiodicity and order}
\label{QCsec:1}

In mathematics, interest in non-periodic tilings of space arose in the
1960s, in the context of the decidability of the question whether a
given finite set of prototiles admits a tiling of the plane \cite{B66}.
Some twenty years later, the unexpected discovery of aperiodically
ordered crystals in intermetallic alloys \cite{SBGC84} brought the
subject to the attention of crystallographers, physicists and
materials scientists. The existence of quasicrystals raises
fundamental questions about the concept of order in nature, and
inspired the ongoing investigation of the associated mathematical
structures.

\subsection{Quasicrystals}
 
Crystals are a paradigm of order in nature. Their symmetry, perfection
and beauty reflects a perfectly ordered structure at the atomic
level. Essentially, the structure of a conventional crystal is based
on a single building block, the unit cell, which usually contains a
small number of atoms, and the entire crystal is then made up by
periodic repetition of the same building block.  For example, in a
salt crystal, the unit cell can be chosen to have cubic shape
containing an equal number of sodium and chlorine atoms, thus giving
rise to a structure with cubic symmetry which is reflected in the
morphology of salt crystals.

\begin{figure}\centering
\includegraphics[width=0.375\textwidth]{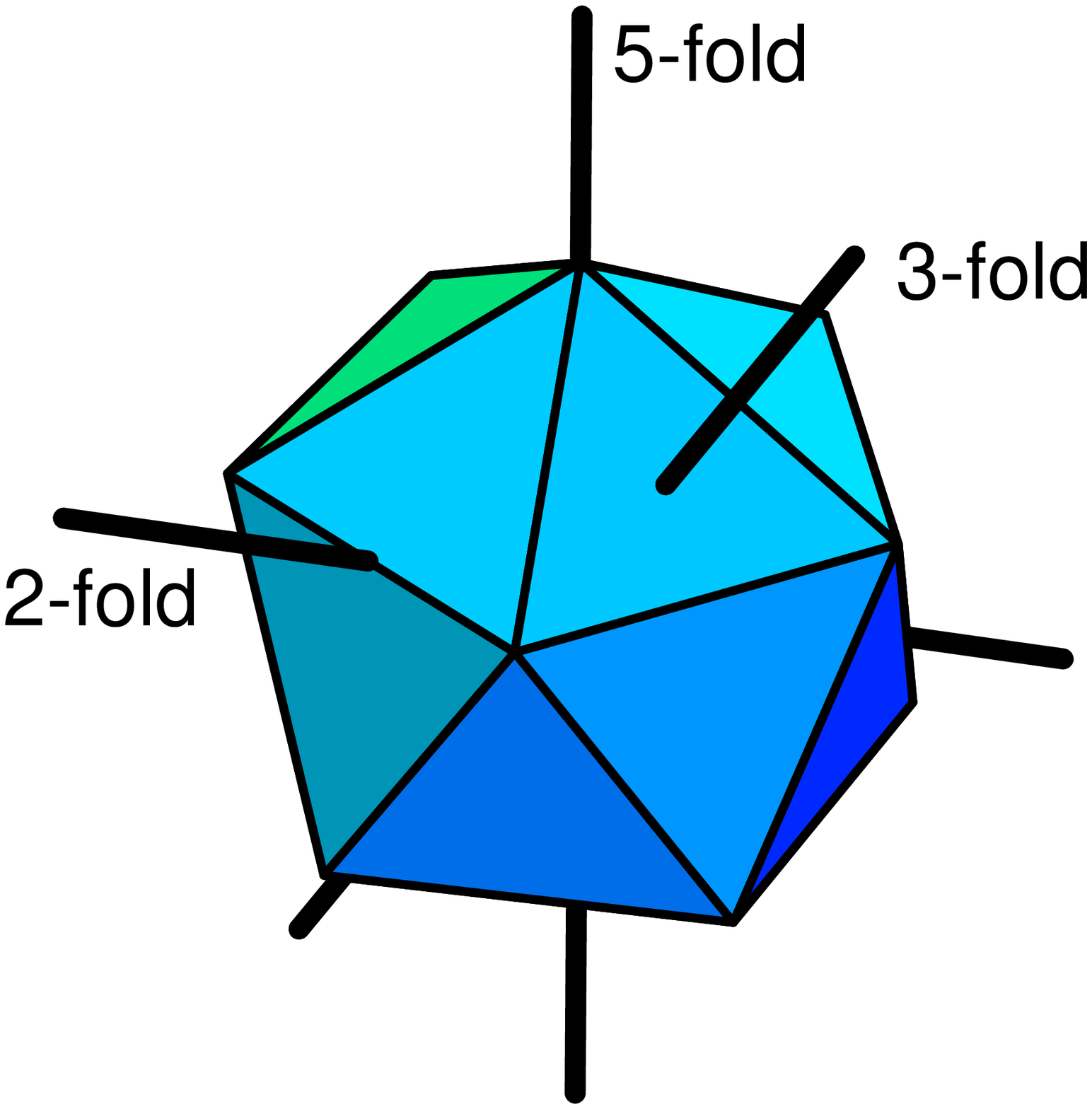}\hspace{0.05\textwidth}
\includegraphics[width=0.375\textwidth]{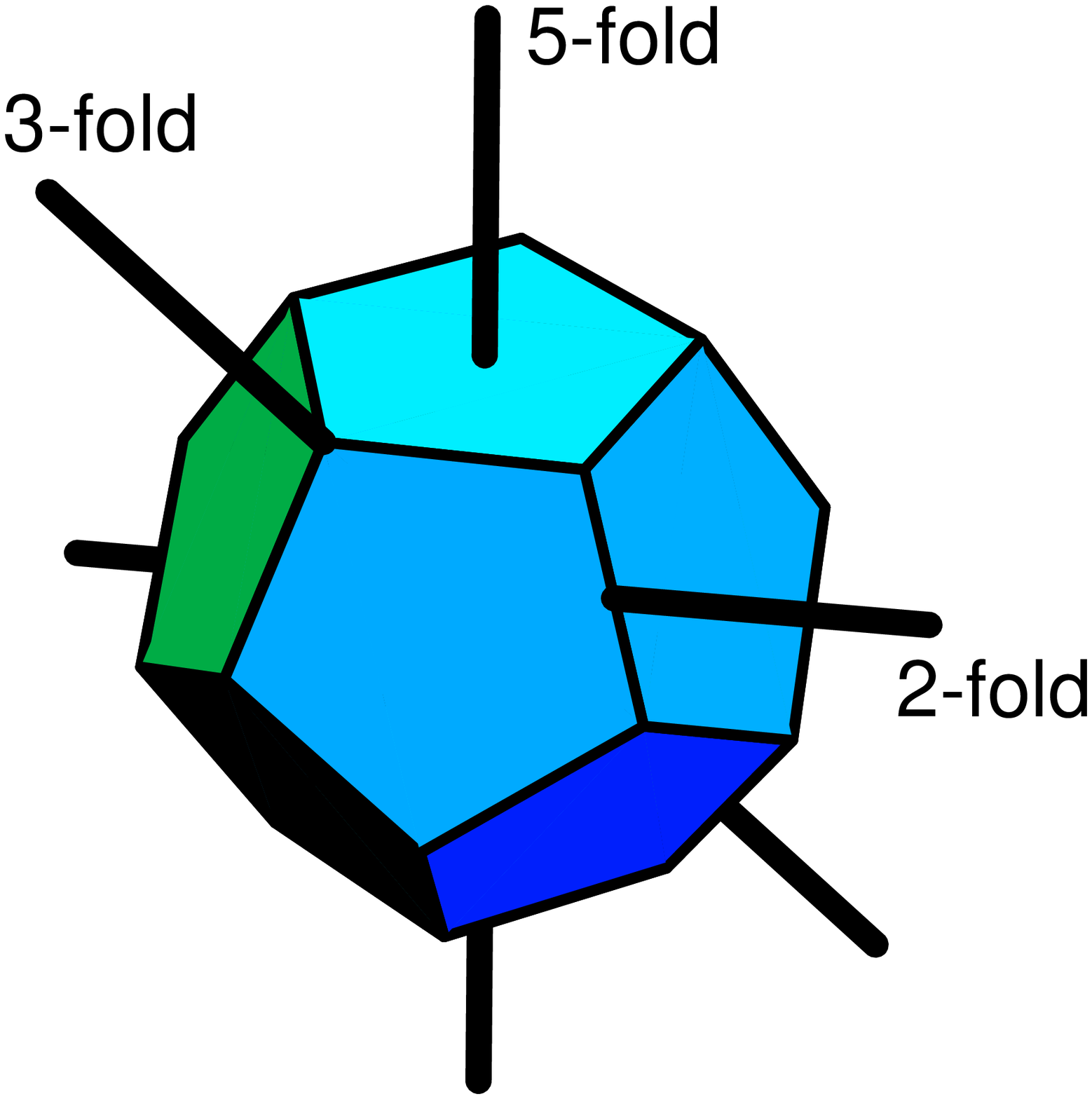}
\caption{The icosahedron (left) and dodecahedron (right). Symmetry
axes of order two, three and five are indicated. There are 15 axes of
order 2 (half the number of edges in both cases), 10 axes of order 3
(half the number of faces in the icosahedron or vertices
in the dodecahedron) and 6 axes of order 5 (half the number of
vertices in the icosahedron or faces in the
dodecahedron).\index{icosahedron}\index{dodecahedron}\label{UG:fig01}}
\end{figure}

Crystalline structures have been classified according to their
symmetries; we shall discuss the corresponding crystallographic point
groups in more detail below. For crystals that are built by periodic
repetition of a single building block, hence are based on a
three-dimensional lattice, the possible symmetries are limited by the
\emph{crystallographic restriction}\index{crystallographic
restriction}. Only certain rotational symmetries are compatible with a
periodic arrangement in three dimensional space; in particular, such
crystals can only have rotational symmetry axes of order $1$, $2$,
$3$, $4$ and $6$. Fivefold or eightfold symmetry, for instance, are
crystallographically `forbidden', and are not found in conventional
crystals. In particular, this includes \emph{icosahedral
symmetry}\index{icosahedral symmetry}, which is the symmetry of the
icosahedron and the dodecahedron, two Platonic solids shown in
Fig.~\ref{UG:fig01}.

\begin{figure}
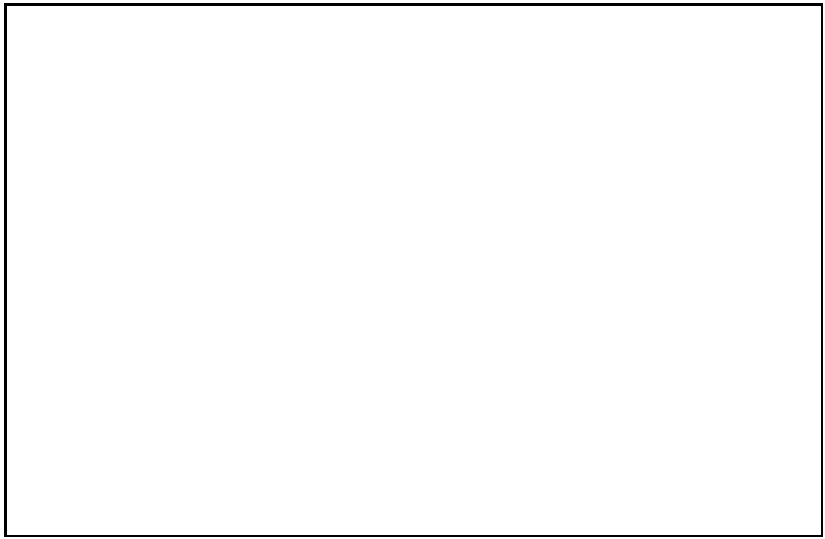
\centering
\framebox{\rule{0pt}{0.58\textwidth}\rule{0.9\textwidth}{0pt}}
\caption{The first published evidence of icosahedral crystals is this
selected area electron diffraction pattern obtained by Dan 
Shechtman \cite{SBGC84} from
a a rapidly cooled aluminium manganese alloy. Angles are measured with 
respect to a
fivefold axis, and the observed 2-, 6- and 10-fold symmetries in the
diffraction patterns match the orientation of 2-, 3- and 5-fold axes
of icosahedral symmetry, compare Fig.~\ref{UG:fig01}. 
Reprinted figure 
with permission from D.~Shechtman, I.~Blech, D.~Gratias, J.W.~Cahn,
Phys.\ Rev.\ Lett.\ \textbf{53}, 1951 (1984).
Copyright (1984) by the American Physical Society.\label{UG:fig02}}
\end{figure}

It thus came as a surprise when in 1982 electron diffraction patterns
of a rapidly cooled aluminium manganese alloy showed a clear account
of icosahedral symmetry \cite{SBGC84}, which includes `forbidden'
five-fold symmetry directions, see Fig.~\ref{UG:fig02} (note that an
$n$-fold symmetric crystal, with odd $n$, produces a $2n$-fold
symmetric diffraction image). Almost simultaneously, twelve-fold
rotational symmetry was observed in a nickel chromium alloy
\cite{INF85}. These materials have the property that their diffraction
patterns\index{diffraction} consist of well defined sharp spots,
similar as observed for conventional crystals, which indicates a
long-range order of the atomic positions. Due to this structural
similarity to conventional crystals,\index{crystal} these new solids
were called \emph{quasicrystals}\index{quasicrystal}. The distinctive
property of quasicrystals is their crystallographically forbidden
symmetry,\index{forbidden symmetry} which can be observed in
diffraction experiments. Such symmetries cannot be accommodated in a
periodic lattice structure in three space dimensions, and hence
quasicrystals cannot be described by a periodic structure in space
based on the repetition of a single unit cell. This means that one is
forced to give up the requirement of periodicity, and indeed
\emph{aperiodic} tilings of space can account for the observed
symmetries, as will be discussed below.

The two examples mentioned above correspond to two different types of
quasicrystals.  The diffraction pattern for the aluminium manganese
quasicrystal found by Shechtman in 1982 displays icosahedral symmetry;
such quasicrystals are known as \emph{icosahedral
quasicrystals}.\index{icosahedral quasicrystal} The arrangement of
atoms in these quasicrystals is not periodic in any direction of
space. In contrast to this, the nickel chromium quasicrystals
discovered by Ishimasa and Nissen have a single direction of
twelve-fold symmetry, and show periodicity along this twelve-fold
symmetry direction. Such systems are called \emph{dodecagonal
quasicrystals},\index{dodecagonal quasicrystals} and you can think of
them as consisting of layers in which atoms are arranged in a
non-periodic, twelve-fold symmetric fashion, which are then stacked
periodically in space. Subsequently, such layered quasicrystals have
also been found with ten-fold (\emph{decagonal
quasicrystals})\index{decagonal quasicrystal} \cite{Bend85} and
eight-fold (\emph{octagonal quasicrystals})\index{octagonal
quasicrystal} \cite{WCK87} symmetry. Over the past twenty years, many
other alloy compositions have been shown to give rise to
quasicrystals, in particular icosahedral and decagonal phases, but no
further symmetries have been found as yet.

\begin{figure}
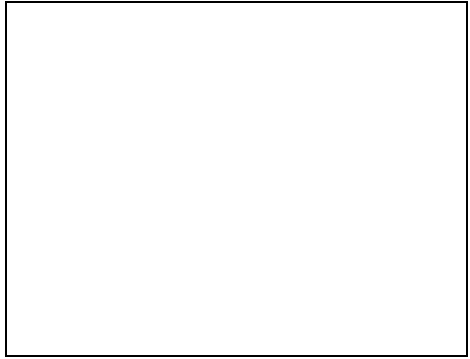
\centering
\framebox{\rule{0pt}{0.38\textwidth}\rule{0.5\textwidth}{0pt}}
\caption{A holmium magnesium zinc `single' quasicrystal \cite{FCPCCOD99}. 
It shows perfect dodecagonal morphology, 
compare Fig.~\ref{UG:fig01}. The background shows a millimetre scale. 
Reprinted figure with permission from 
I.R.~Fisher, K.O.~Cheon, A.F.~Panchula, P.C.~Canfield,
M.~Chernikov, H.R.~Ott, K.~Dennis, Phys.\ Rev.\ B \textbf{59}, 
308 (1999).
Copyright (1999) by the American Physical Society.\label{UG:fig03}}
\end{figure}

As for a conventional crystal, the symmetry of the atomic arrangement
in a quasicrystal often manifests itself in the morphology of
high-quality specimens. Detailed and laborious investigations of phase
diagrams of ternary alloy systems has made it possible to grow
`single' quasicrystals from the melt, providing well characterised
samples for experimental studies. An example of a dodecahedral crystal
of an holmium magnesium zinc quasicrystal is shown in
Fig.~\ref{UG:fig03}, showing beautiful facets that perfectly reflect
the intrinsic icosahedral order of the alloy. As a consequence of
their peculiar atomic arrangements, quasicrystals show interesting
physical properties, leading to a number of promising practical
applications, in particular as low-friction surface coatings and
as storage material for hydrogen. 

The aperiodic tilings discussed below offer a simple model structure
than can explain the observed symmetries in quasicrystals, and serve
as appropriate mathematical idealisations of the atomic structure, in
analogy to the role of lattices in conventional crystallography. In
nature, there is no truly perfect crystal (even the most expensive
diamond contains some defects), and in the same sense the structure of
a real quasicrystal will differ considerably from these idealised
tilings.  In particular, many quasicrystals are high-temperature
phases, which hints at the importance of entropy for the stability of
these phases, in which case you would expect an inherently disordered
structure. Like many other questions concerning quasicrystals, this is
subject of current research activity and scientific debate, and has
not yet been satisfactorily understood; see also \cite{JA,SE} and
\cite{SSH02,TR} for introductory monographs and collections of
introductory articles on the mathematics and physics of quasicrystals,
and \cite{MO,BM,KP} for more in-depth mathematical results.

\subsection{Crystallographic restriction}
\index{crystallographic restriction}

\begin{figure}\centering
\includegraphics[width=0.9\textwidth]{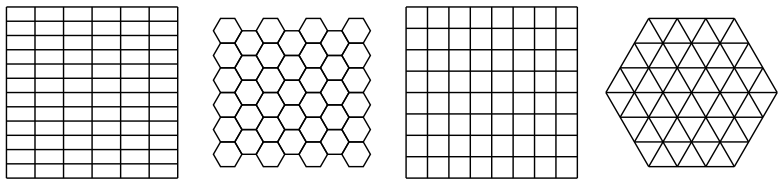}
\caption{Planar periodic tilings with 2-, 3-, 4- and 6-fold rotational 
symmetry.\label{UG:fig04}}
\end{figure}

Before discussing examples of such tilings, we briefly explain the
\emph{crystallographic restriction}\/ mentioned above. It states that
in a periodic lattice in two or three dimensions, the only possible
non-trivial rotation symmetries are 2-, 3-, 4- and 6-fold symmetry.
It is easy to come up with examples of periodic structures that have
these symmetries, see Fig.~\ref{UG:fig04}. How can one see that other
symmetries, such as fivefold symmetry, cannot occur?\index{rotational
symmetry}\index{fivefold symmetry}

A simple argument goes as follows. Assume for a moment that you had a
periodic lattice with fivefold symmetry. This means that you can
rotate the lattice by multiples of $72^{\circ}$ about any of its lattice
points, and obtain the same lattice again. We shall now show that this
is impossible.\index{lattice}\index{periodicity}

\begin{figure}\centering
\includegraphics[width=0.58\textwidth]{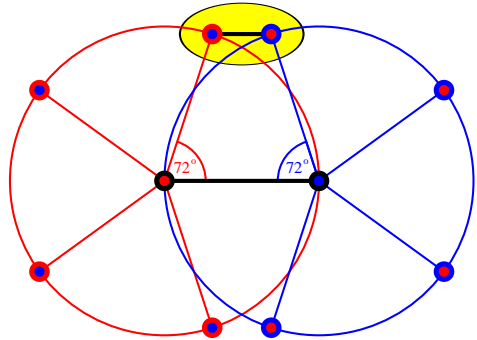}
\caption{Two starting points (dark), and the two sets of 4 points
obtained by rotating one about the other by multiples of
$72^\circ$. The highlighted pair of rotated points is closer than the
pair of rotation centres, leading to the
contradiction.\label{UG:fig05}}
\end{figure}

Start with two lattice points which have minimal distance from each
other.  Then, we can rotate one point about the other by multiples of
$72^{\circ}$, and the four new points obtained in this way must be
lattice points as well. Analogously, we can rotate choosing the other
point as a centre, which gives us another four new lattice points, see
Fig.~\ref{UG:fig05}. But now you see that there are lattice points
that are closer to each other than the ones we started from, which
were supposed to have minimal distance -- so we end up with a
contradiction, which means that there is no such structure.

\begin{figure}\centering
\includegraphics[width=0.4\textwidth]{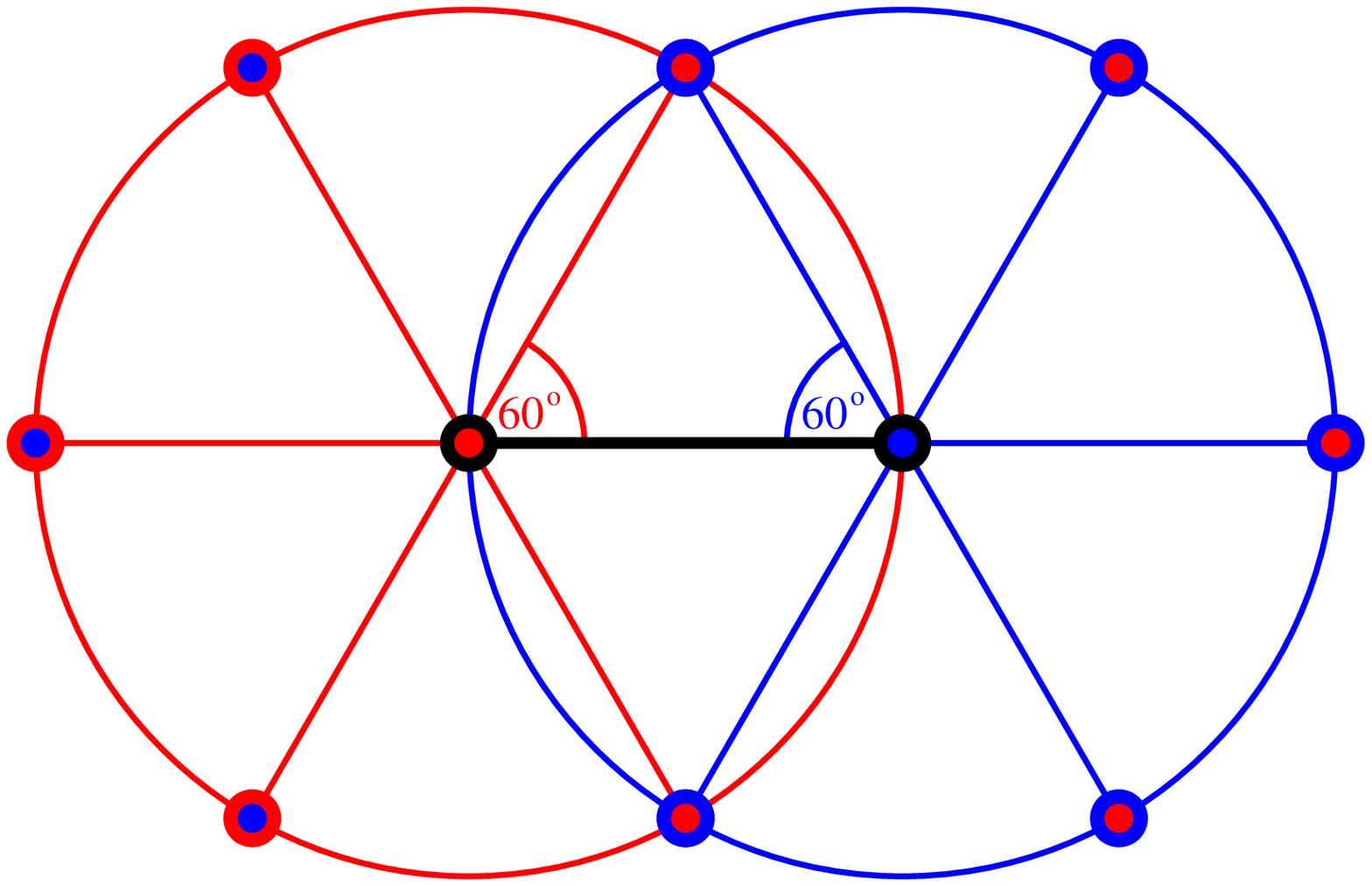}\quad
\includegraphics[width=0.4\textwidth]{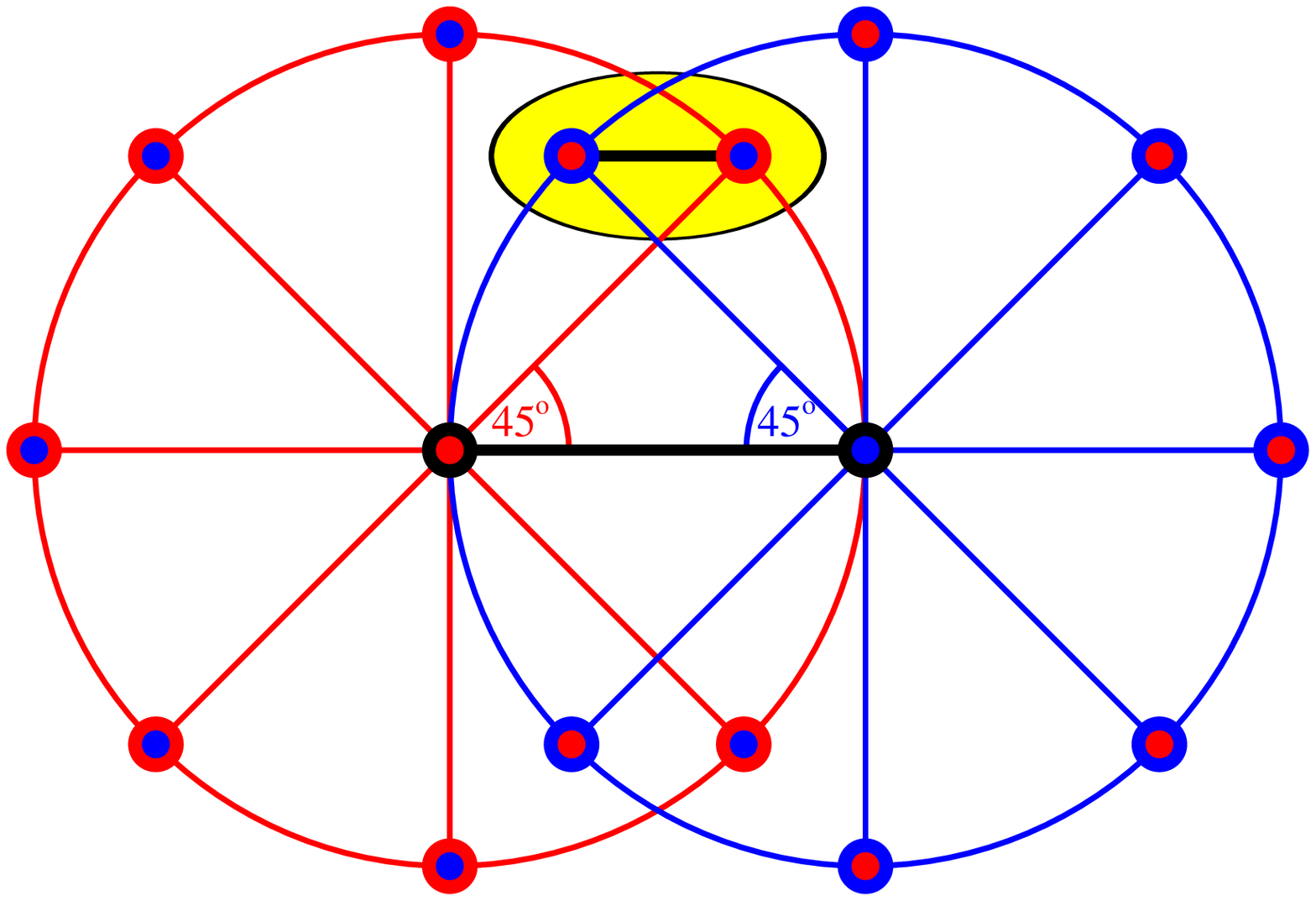}
\caption{The same as Fig.~\ref{UG:fig05} for the 6-fold (left) and
8-fold (right) case.  For the 6-fold case, the points obtained by
rotation form part of a triangular lattice, 
whereas the 8-fold case again leads to a
contradiction. Obviously, the same happens for any rotation angle
$360^{\circ}/n$ with $n>6$, since the corresponding 
highlighted pair of rotated
points is always closer together 
than the two original points.\label{UG:fig06}}
\end{figure}

If you use the same argument for 2-, 3-, 4- and 6-fold symmetry, no
problem arises. Fig.~\ref{UG:fig06} shows (on the left) the case of
sixfold symmetry, where rotated latticed points coincide to form an
equilateral triangle, which can be extended to a triangular lattice
shown in Fig.~\ref{UG:fig04} on the right. For any $n$-fold rotational
symmetry with $n>6$, this will not happen, and you always obtain two
lattice points that are closer than the points you started with, see
the right part of Fig.~\ref{UG:fig06} for an example with $n=8$.

No new possibilities arise in three dimensions; however, this is not
true for dimensions larger than three. In particular, periodic
lattices with 5-, 8- and 12-fold symmetry exist in four dimensions,
and in six dimensions you find lattices with full icosahedral
symmetry.\index{icosahedral symmetry}

\subsection{Aperiodic tilings}
\index{aperiodic tiling}

To overcome the limitations imposed by the crystallographic
restriction, we have to look at more general classes of tilings,
without lattice periodicity. As discussed below, aperiodically ordered
tilings can be constructed that reproduce the symmetry seen in
diffraction experiments. Paradigmatic examples of such structures are
the \emph{Penrose tiling} \cite{PE} and the \emph{Ammann-Beenker
tiling} \cite{AGS92}.\index{Penrose tiling}\index{Ammann-Beenker tiling}
In the following, we briefly introduce these tilings and explain how
they can be constructed, while heuristically motivating some of their
properties along the way; see also \cite{BGM02} for a gentle
introduction, and \cite{GS02} for the computer generation of the
tilings shown below. A deeper description of the underlying
mathematics is given in Sect.~\ref{QCsec:2}.

We mention three rather different approaches -- matching rules,
inflation and projection from higher-dimensional periodic
lattices. While the usual examples can be described in any of these
settings, they are by no means equivalent. For instance, there are
many non-periodic tilings that can be obtained by inflation, but cannot
be embedded in a periodic lattice in a finite-dimensional
space. Arguably the most powerful approach is the projection method;
for example, it can be shown in a rather general setting that such
point sets are pure point diffractive, which means that they give rise
to point-like diffraction patterns\index{diffraction} 
such as shown in Fig.~\ref{UG:fig02}.

\subsubsection{Matching rules}
\index{matching rule}

\begin{figure}\centering
\includegraphics[width=0.5\textwidth]{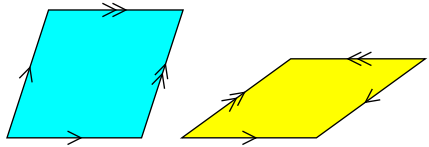}
\caption{The tiles of the rhombic Penrose tiling with
arrow decorations.\label{UG:fig07}}
\end{figure}

Seemingly the simplest way to specify an aperiodic tiling, such as the
Penrose tiling, is by so-called \emph{matching rules}. In essence,
these are specific rules that restrict the possible local arrangements
of the basic tiles. In the simplest examples, matching rules can be
encoded by markings (or \emph{decorations}\index{decoration}\/) on,
for instance, the edges of the tiles. An example is the rhombic
Penrose pattern, which is obtained from two rhombic prototiles with
single and double arrows on the edges, as shown in
Fig.~\ref{UG:fig07}.\index{Penrose tiling}\index{marked tiles}

\begin{figure}\centering
\includegraphics[width=0.9\textwidth]{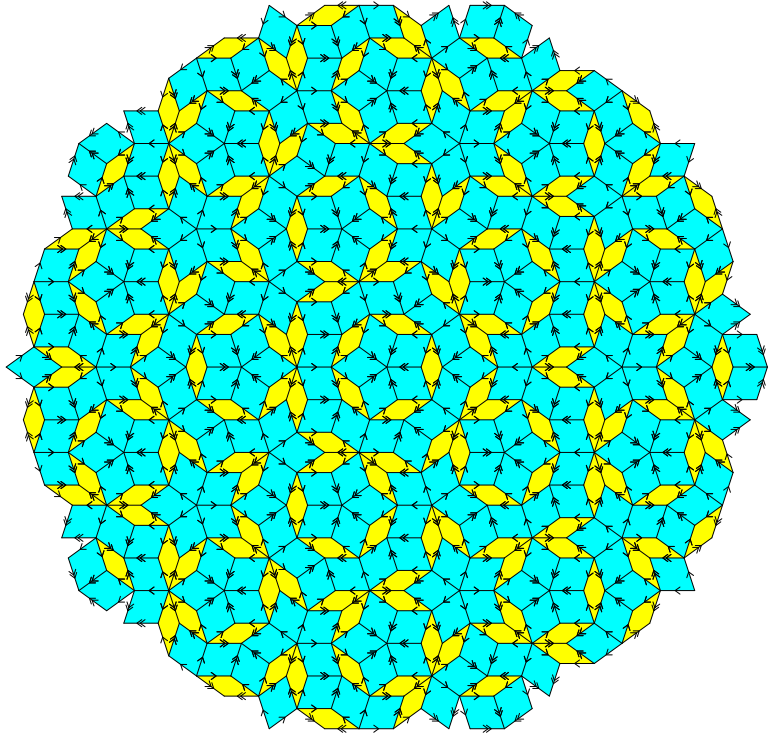}
\caption{A legal patch of a Penrose tiling.\label{UG:fig08}}
\end{figure}

Clearly, ignoring the arrows these tiles can give rise to periodic
tilings of the plane -- as an example just take the tiling made by
repeating one of the two tiles periodically. To obtain the Penrose tiling,
tiles are assembled subject to the constraint that tiles in the tiling
are edge-to-edge and such that the arrow decorations on adjacent edges
match. These matching rules `enforce' aperiodicity. A `legal' patch is
shown in Fig.~\ref{UG:fig08}.\index{legal patch}\index{aperiodicity}

Note that the matching rules do \emph{not}\/ amount to an algorithm
that allows you to grow a Penrose tiling. Indeed, it is more like a
jigsaw puzzle -- if you start putting tiles together you may arrive at
a situation where neither of the tiles fits at a particular place, in
which case the patch you have grown does not occur in a perfect
Penrose tiling. However, if you manage to grow a tiling that fills the
plane it is indeed `the' Penrose tiling \cite{PE} (or more precisely,
a member of the class of \emph{locally indistinguishable}\/ Penrose
tilings).\index{local indistinguishability} In the application to
quasicrystals, where it is tempting to associate matching rules to
preferred local arrangement of atoms, this has sparked discussions
about how quasicrystals grow, a topic that is still not completely
understood, see \cite{GJ02} for a review.

\begin{figure}\centering
\includegraphics[width=0.95\textwidth]{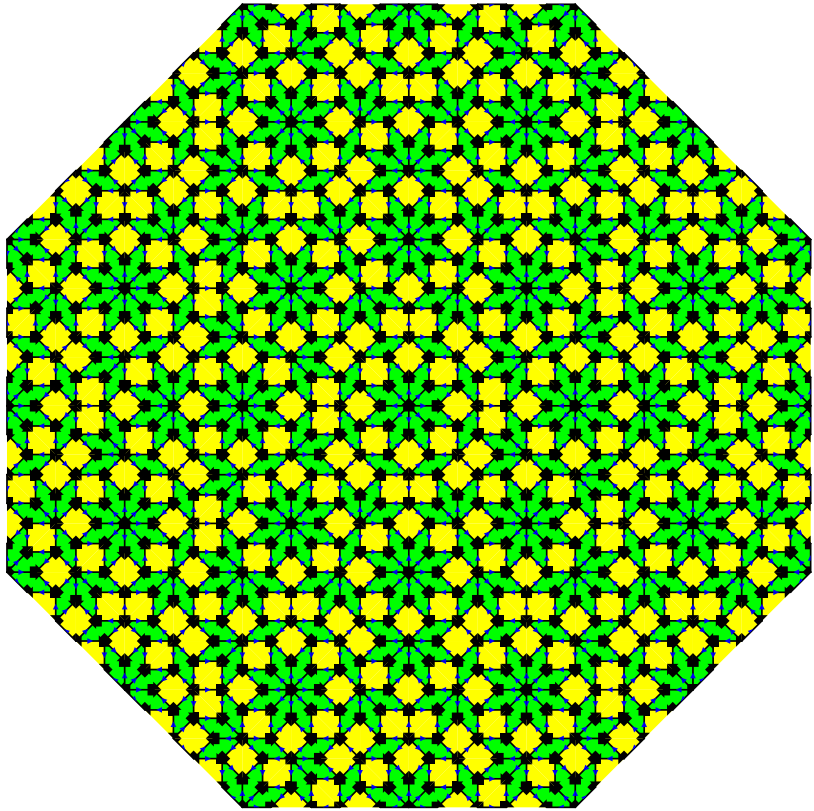}
\caption{A legal patch of an Ammann-Beenker tiling.\label{UG:fig09}}
\end{figure}

The Ammann-Beenker tiling, again built from two different tiles, has
eightfold rotational symmetry.\index{eightfold symmetry} It also
possesses matching rules; however, in this case decorations on edges and
vertices are needed to exclude periodic tilings. A legal patch of the
tiling, including the matching rules, is shown in
Fig.~\ref{UG:fig09}.\index{Ammann-Beenker tiling}

\subsubsection{Inflation}
\index{inflation}

\begin{figure}\centering
\includegraphics[width=0.6\textwidth]{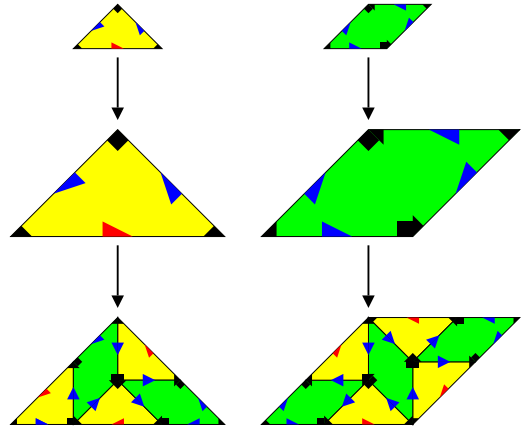}
\caption{Inflation of the Ammann-Beenker tiling.\label{UG:fig10}}
\end{figure}

An important concept for the construction of aperiodically ordered
structures is the so-called inflation (or deflation) procedure. It is
based on a transformation of the tiles which consists of two steps --
the re-scaling by a constant factor (\emph{inflation
factor})\index{inflation factor} and the dissection of the tiles into
a number of copies of the original tiles, of the original size. An
inflation rule for the Ammann-Beenker tiling is shown in
Fig.~\ref{UG:fig10}, where for simplicity we use a triangle (half a
square) as the basic tile. For this case, the inflation factor is
$1+\sqrt{2}$. Note that the dissection of the triangle breaks the
reflection symmetry of the unmarked tiles -- you need to consider the
orientation of the triangle, marked by the arrow on the
hypotenuse. The dissection of a triangle of the opposite orientation
is the mirror image of the one shown here.

Now, you can build an Ammann-Beenker tiling by repeated application of
this rule, starting for instance from a single tile. After one step,
you obtain one of the patches shown at the bottom of
Fig.~\ref{UG:fig10}. Applying the procedure again, you obtain a patch
consisting of more tiles, and so on and so forth. In this way, you can
create an arbitrarily large tiling, and the infinite structure
obtained in the limit has the property that is is invariant under the
inflation procedure. In fact, the octagonal patch of the
Ammann-Beenker tiling depicted in Fig.~\ref{UG:fig09} was obtained
precisely in this way, by applying inflation starting with an
eightfold symmetric patch.

The inflation symmetry\index{inflation symmetry} of an infinite
Ammann-Beenker tiling has an interesting consequence. If you cut out any
finite patch from the tiling, then this patch will occur in the
infinite tiling over and over again. This property is called
\emph{repetitivity}\/ (not to be confused with
periodicity).\index{repetitivity} You can deduce this property by
applying the inverse transformation,
\emph{deflation},\index{deflation} to the tiling with your chosen
patch. Eventually, after a finite number of steps, your patch will be
mapped to a single tile. Since the two tiles occur with positive
frequency in the tiling (which can be calculated from the inflation
rule), if follows by applying the same number of inflation steps again
that the chosen patch also occurs with positive frequency.

In this sense, the Ammann-Beenker tiling is very regular indeed --
like in a periodic structure any finite patch repeats in a rather
regular fashion, albeit not periodically. So if you find the same
feature repeated at a certain distance, it will not necessarily repeat
at twice the distance -- it may repeat earlier or later, but it will
repeat.

\subsubsection{Cut and project sets}
\index{cut and project set}

\begin{figure}\centering
\includegraphics[width=0.24\textwidth]{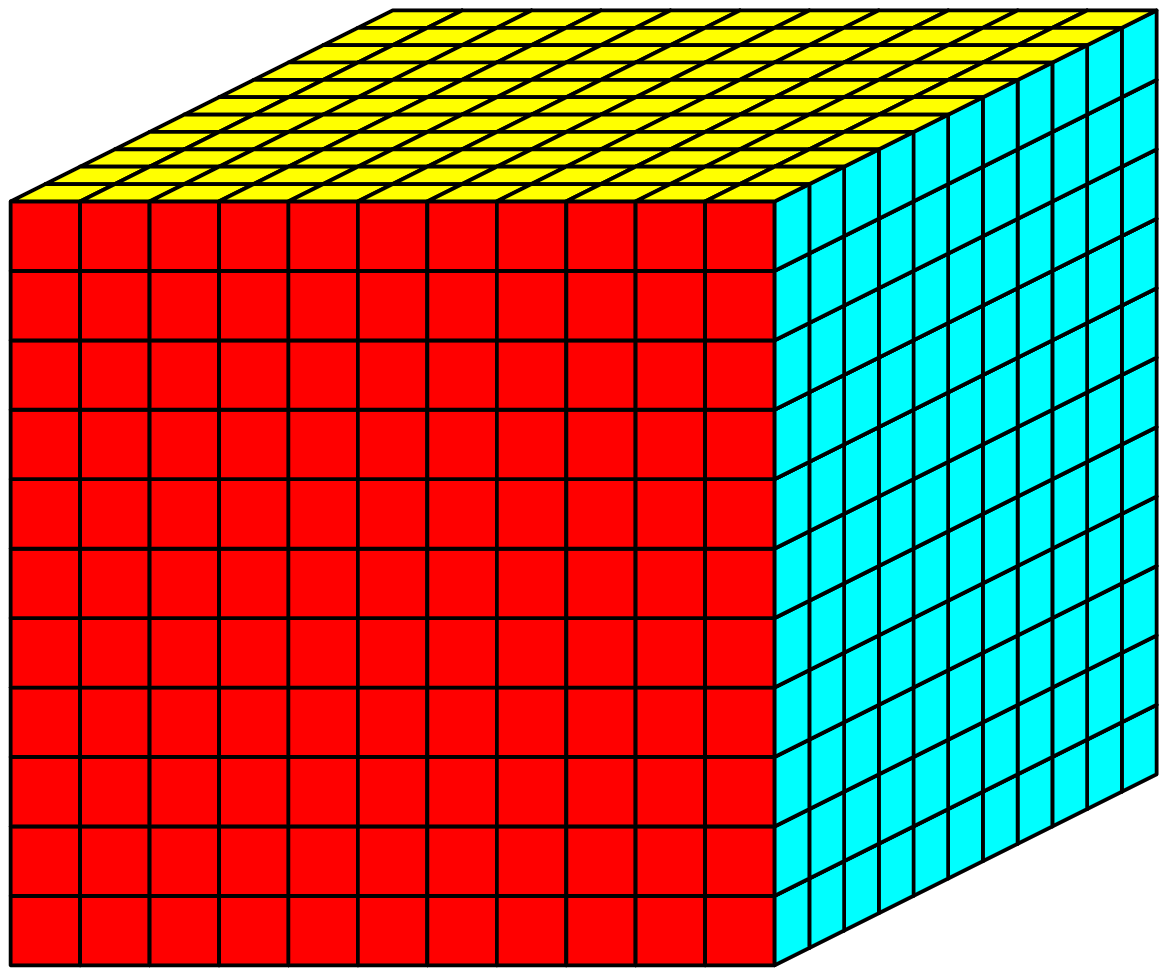}
\includegraphics[width=0.24\textwidth]{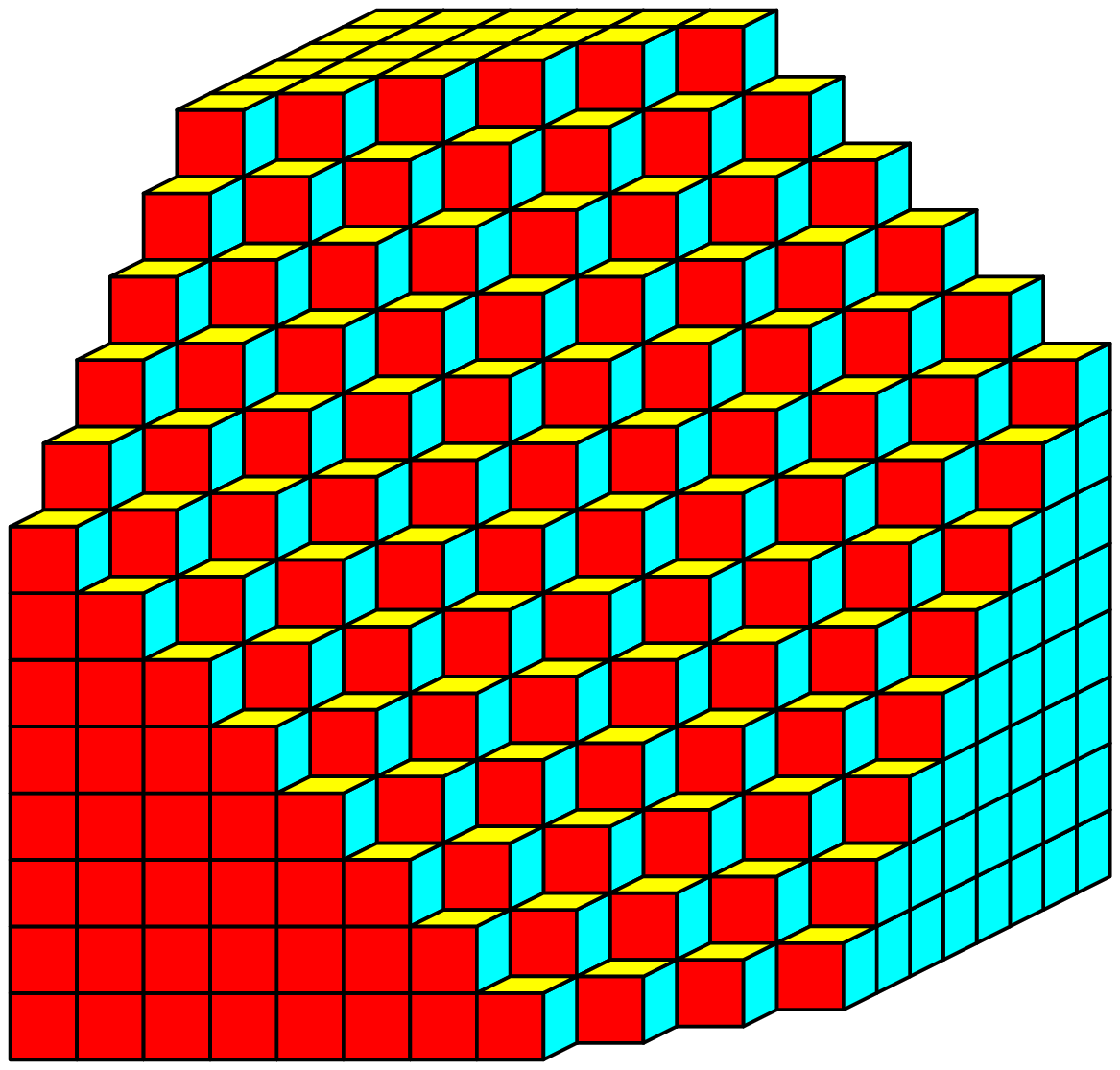}
\includegraphics[width=0.40\textwidth]{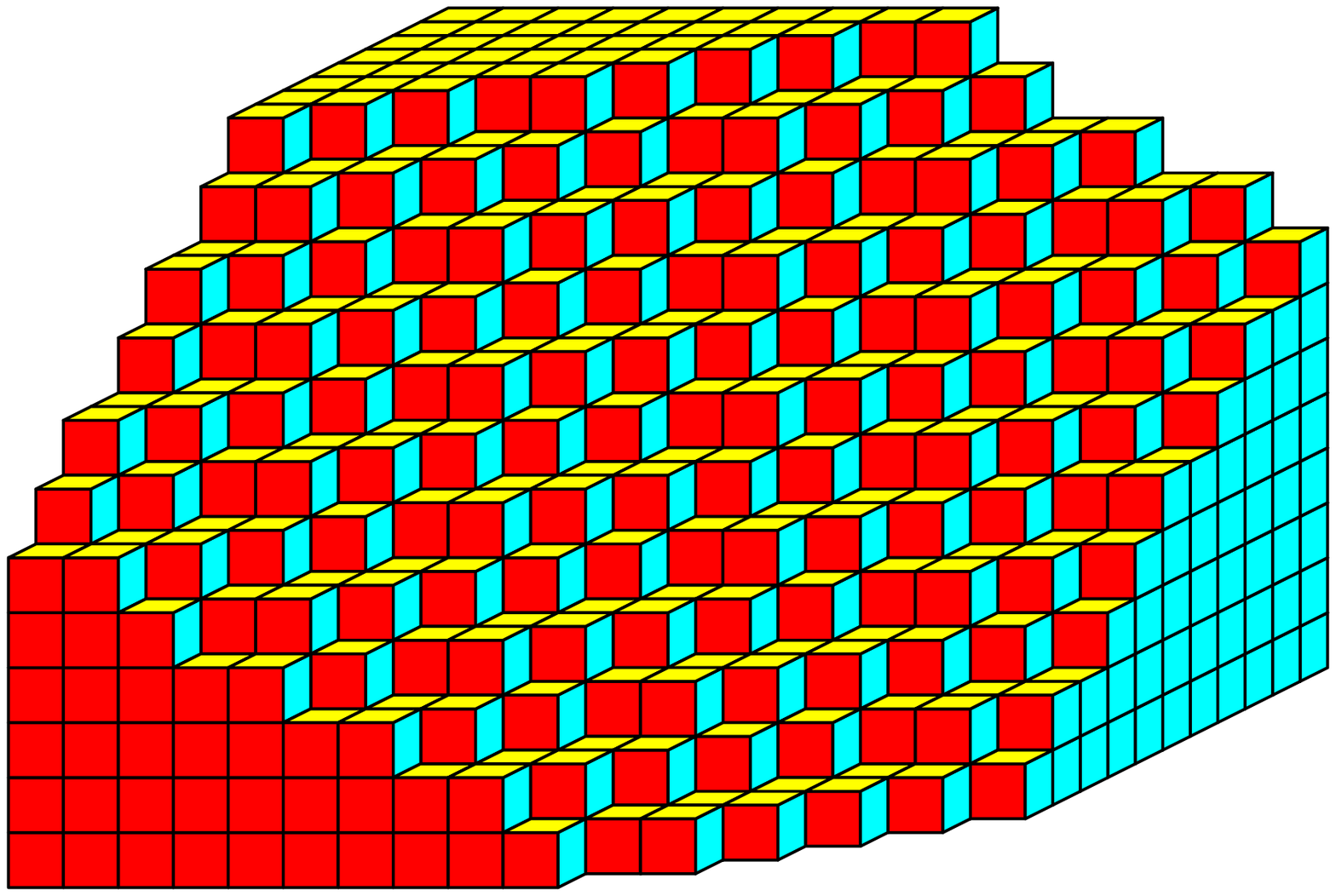}
\caption{Planar projections of a packing of cubes, showing different
surfaces. The projected planar tiling consist of three rhombic tiles,
corresponding to the projections of the three visible faces of the
cube.  Onlye for special surfaces, such as shown on the left and in
the centre, the rhombs form a periodic tiling; in general, the rhombic pattern 
on a surface is non-periodic.\label{UG:fig11}}
\end{figure}

The third, and final method discussed in this section, is based on
periodic lattices, but in a higher-dimensional space. Let us consider
a simple example first. Imagine looking at a huge stack of sugar
cubes, all nicely arranged in a regular fashion to form a cubic
structure in three dimensions. If you take a horizontal surface (so
your pile is a big cube made up of lots of small cubes), and look at
it from some angle, you will see a pattern consisting of rhombs, as
shown in Fig.~\ref{UG:fig11} on the left. If you look at an inclined
surface at commensurate angles, such as shown in the centre of
Fig.~\ref{UG:fig11}, you find periodic patterns made up of three
different rhombs, which are the projections of the faces of cubes, and
their shape depends on the inclination of the surface. If you look at
a more general surface, you still find a tiling made up of the same
three rhombs. However, the rhombic pattern you observe will, except
for special choices of surface, not be periodic. An example is shown
on the right of Fig.~\ref{UG:fig11}.  Even if the surface is `flat',
as it is in this case, in general you end up with a non-periodic
pattern, such as shown on the right of Fig.~\ref{UG:fig11}.

This is the main idea behind the cut and project scheme. Starting from
a higher-dimensional periodic lattice, you can obtain an aperiodic
tiling by considering the projection of a `slice' of that lattice onto
the \emph{physical space}\index{physical space} of two or three
dimensions. In our example above, we obtained a two-dimensional
aperiodic tiling as a projection from a three-dimensional cubic
lattice.\index{projection}

As it turns out, the Penrose and Ammann-Beenker tilings\index{Penrose
tiling}\index{Ammann-Beenker tiling} both can be described in this
way. However, you need to employ four-dimensional lattices to do
this. In four dimensions, it is possible to have symmetry axes of
order 5 or 8 in a periodic lattice, and indeed the Penrose and
Ammann-Beenker tilings are planar projections of appropriate
`slices' of the corresponding four-dimensional lattices, along the
high-symmetry direction, such that the resulting tiling inherits the
rotational symmetry.\index{rotational symmetry}\index{four-dimensional
lattice}

\begin{figure}\centering
\includegraphics[width=0.7\textwidth]{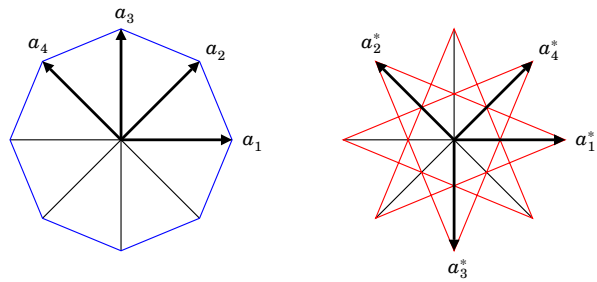}\quad
\caption{The vectors $a_{k}$, $k=1,\ldots 4$ spanning the 
$\mathbb{Z}$-module $\Lambda$ in `physical' space, 
and the corresponding vectors $a_{k}^{*}$ spanning
the `internal' space.\label{UG:fig12}}
\end{figure}

\begin{figure}\centering
\includegraphics[height=9cm]{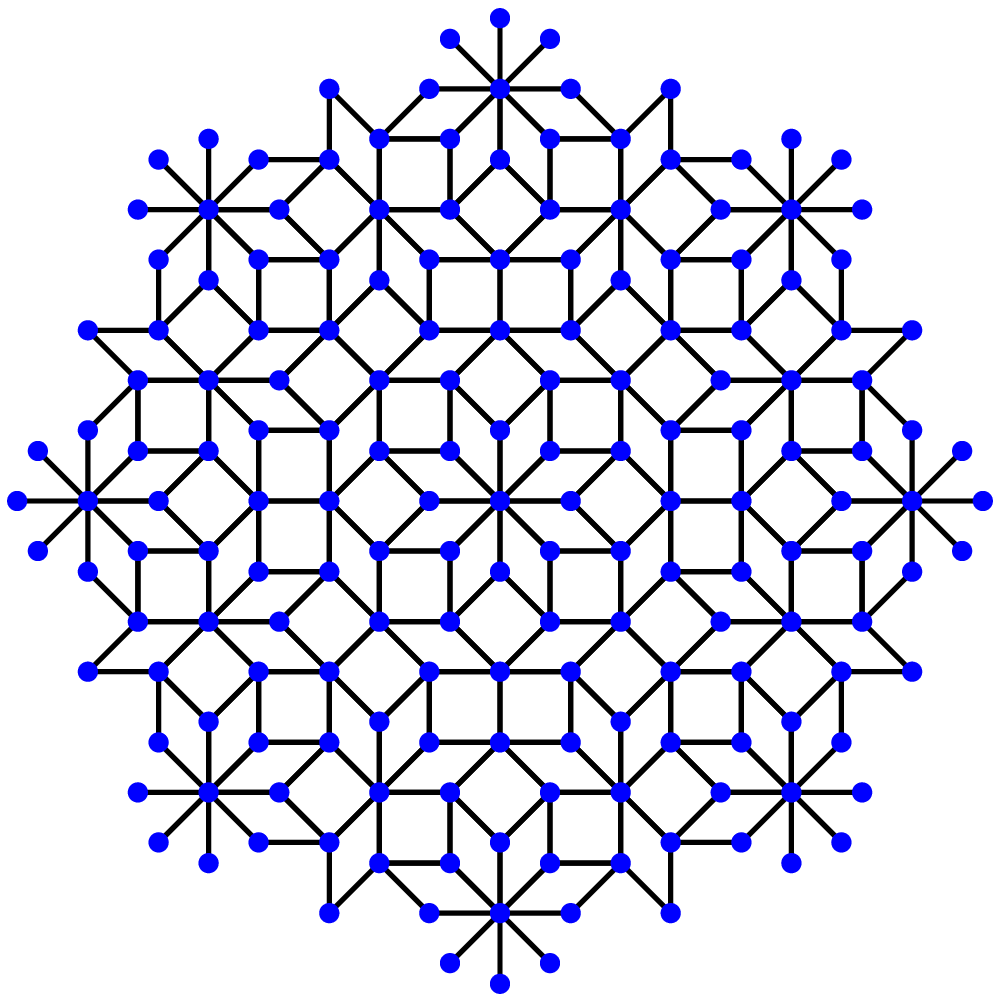}\qquad
\includegraphics[height=9cm]{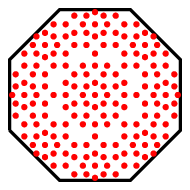}
\caption{The Ammann-Beenker tiling as a cut and project set. On the
left, the projected points $V_{\mathrm{AB}}$ 
of the hypercubic lattice in physical space
are shown, with lines connecting points of unit distance. On the
right, the corresponding projections $V_{\mathrm{AB}}^{*}$ 
in internal space are shown,
which fall into a regular octagon of unit edge length.\label{UG:fig13}}
\end{figure}

\begin{figure}\centering
\includegraphics[width=0.8\textwidth]{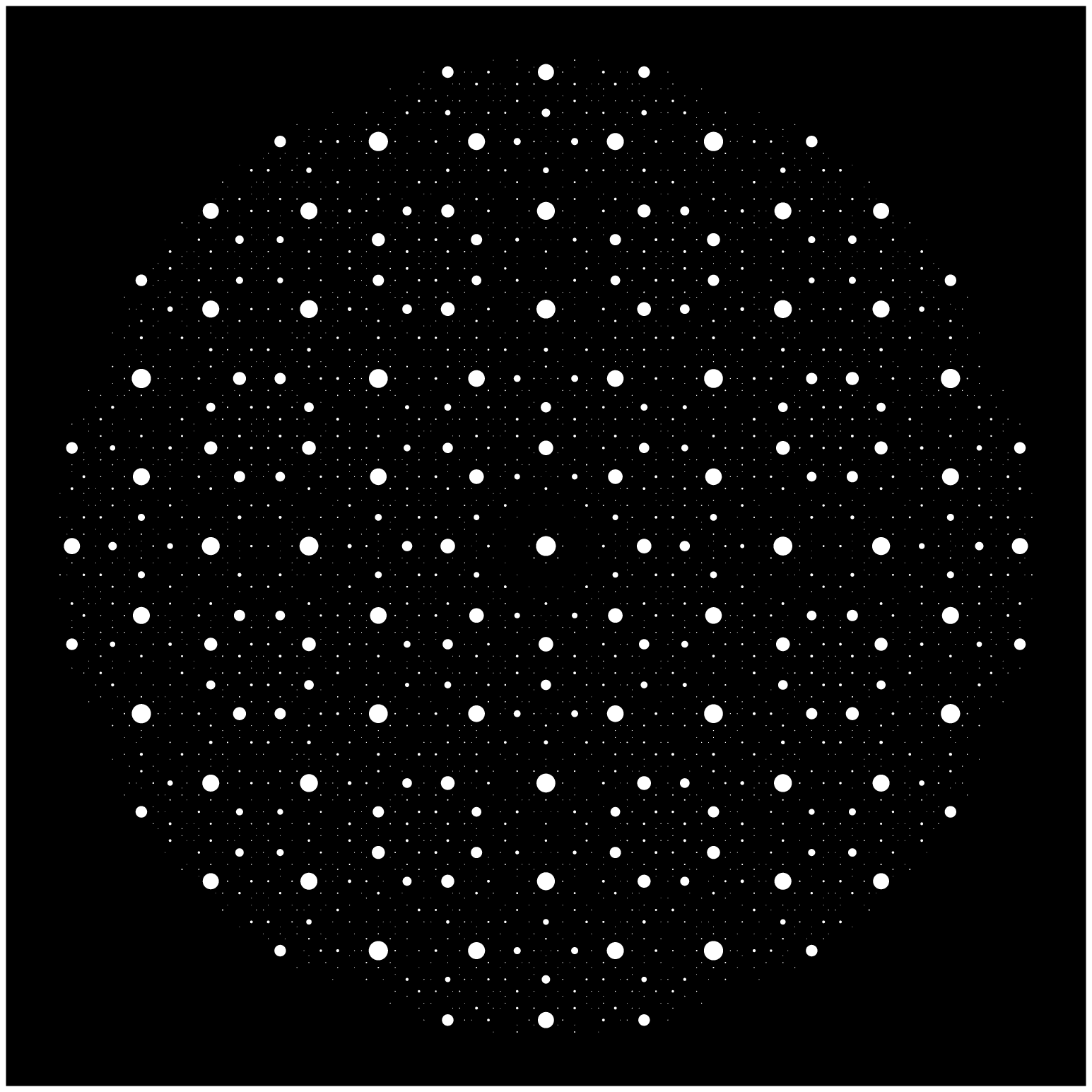}
\caption{Calculated diffraction image of the
Ammann-Beenker tiling.\index{diffraction}\index{Ammann-Beenker tiling}\label{UG:fig14}}
\end{figure}

Explicitly, for the Ammann-Beenker tiling the cut and project approach
can be implemented as follows. Consider the $\mathbb{Z}$-module of all
integer linear combinations of the four vectors $a_{k}$,
$k=1,\ldots,4$, shown in Fig.~\ref{UG:fig12} on the left,
\[
\Lambda = \{n_{1} a_{1} + n_{2} a_{2} + n_{3} a_{3} + n_{4} a_{4} \mid 
(n_{1},n_{2},n_{3},n_{4})\in\mathbb{Z}^{4}\}
\]
which gives you a point set that is dense in the plane, corresponding
to the projection of the entire hypercubic lattice $\mathbb{Z}^4$ onto
the `physical space'. From these, select all points for which the
corresponding integer linear combination $n_{1} a^{*}_{1} + n_{2}
a^{*}_{2} + n_{3} a^{*}_{3} + n_{4} a^{*}_{4}$ of the vectors
$a^{*}_{k}$, shown on the right in Fig.~\ref{UG:fig12}, falls into a
regular octagon $O$ of unit edge length. The vectors $a^{*}_{k}$ span
the projection of the hypercubic lattice $\mathbb{Z}^4$ in the
\emph{internal space}\index{internal space}. The octagon defines the
`slice' of the hypercubic lattice\index{hypercubic lattice} that is
projected, and is called the \emph{window}\index{window}\/ or
\emph{acceptance domain}\index{acceptance domain} of the cut and
project scheme.  Explicitly, the set of vertex points of the
Ammann-Beenker tiling is given by
\[
V_{\mathrm{AB}}=\{x\in\Lambda \mid x^{*}\in O\},
\]
and an example is shown in Fig.~\ref{UG:fig13}.

It can be shown that, under rather general assumptions, cut and
project sets defined in this way possess a pure point diffraction
measure.\index{pure point diffraction} As an example, the calculated
diffraction pattern of scatters positioned on the vertices of an
Ammann-Beenker tiling is shown in Fig.~\ref{UG:fig14}.

\section{Quasiperiodic functions, tilings and coverings}
\index{tiling}\label{QCsec:2}

We proceed to a closer look at the mathematics. Periodic order as
found in Euclidean space $E^3$ is analysed in terms of a lattice\index{lattice} 
$\Lambda$.  Quasiperiodicity was viewed by H.\ Bohr on sections
irrational with respect to a lattice $\Lambda \in E^n$.  The study of
quasicrystals showed how irrational sections emerge from forbidden
point symmetry. The standard cell structure of periodic functions is
absent in quasiperiodicity.  The geometry of the lattice $\Lambda \in
E^n$ provides dual periodic Voronoi and Delone cell complexes.  By the
projection of boundaries from the dual complexes, a canonical tiling
theory for quasiperiodic structures can be found and put in action.

\subsection{Irrational subspaces and quasiperiodicity}

We describe the geometry underlying H.~Bohr's description \cite{BO} of
quasiperiodicity.

Bohr (1925) considers an Euclidean space $E^n$ with scalar product
$\langle, \rangle$ of dimension $n$ and a periodic lattice $\Lambda
\in E^n$.  We first describe rational subspaces with respect to
$\Lambda \in E^n$.  Given the basis vectors $(a_1,a_2,\ldots, a_n)$ of
$\Lambda$, the lattice points are determined by
\begin{equation}
\label{Q1}
\{ t \mid 
t=\sum_j m_ja_j,\; (m_1,m_2,\ldots,m_n)\in \mathbb{Z}^n \}.
\end{equation}  
The reciprocal lattice $\Lambda^R$ has the basis $(b_1,b_2,\ldots,
b_n)$, $\langle b_i, a_j\rangle =\delta_{ij}$.  Any fixed reciprocal
lattice vector $ b \in \Lambda^R$, determines a rational hyperplane
$Y_b$ of the lattice $\Lambda$ with points characterised by
\begin{equation}
\label{Q2}
Y_b= \{ x \mid \langle x, b \rangle =0\}.
\end{equation}  
This rational hyperplane is parallel to a hyperplane containing 
the lattice points 
\begin{equation}
\label{Q3}
\{ t \mid \langle t, b \rangle =0\}.
\end{equation} 
A general rational linear subspace is the intersection of rational
hyperplanes.  Any other linear subspace for $\Lambda \in E^n$ is
called an irrational subspace with respect to $\Lambda \in E^n$.

\begin{example}
The irrational Fibonacci subspace of $E^2$ is a line 
of slope $\tau=\frac{1}{2}(1+\sqrt{5})$ through the square lattice
$\Lambda =\mathbb{Z}^2$. It is a horizontal line in Fig.~\ref{PK:fig2}.
Below we shall construct the Fibonacci tiling on this
irrational line. 
\end{example}

A periodic function\index{periodic function}
$f^p(x),\; x \in E^n$ fulfils for all lattice translations $t \in \Lambda$
\begin{equation}
\label{Q3a}
f^p:\; f^p(x+t)=f^p(x).
\end{equation}
Now split $E^n$ into two orthogonal complementary linear subspaces
\begin{equation}
\label{Q4}
E= E_{\parallel} + E_{\perp},\quad E_{\parallel}\; \perp\;  E_{\perp},
\end{equation}
such that 
\begin{itemize}
\item[(i)]$E_{\parallel}$ is a subspace $E_{\parallel}=E^m<E^n$ of 
dimension $m<n$ and 
\item[(ii)] $E_{\parallel}$ is irrational with respect to 
$\Lambda \in E^n$. 
\end{itemize}

Define the Bohr class of  quasiperiodic functions\index{quasiperiodic function} 
$f^{qp}(x_{\parallel})$
as the restrictions of  $\Lambda$-periodic functions $f^p$ from $E^n$ to $E^m$,
\begin{equation}
\label{Q5}
f^{qp}(x_{\parallel}):= 
f^p(x)|_{x= x_{\parallel}+c_{\perp}}, \quad c_{\perp}= {\rm const.}
\end{equation}
Recall that the periodic function $f^p$ has a pure point Fourier
spectrum.  By transferring the restriction of Eq.~(\ref{Q5}) to Fourier
space, the quasiperiodic function $f^{qp}$ can be shown to also
possess a pure point spectrum. It is carried by a countable but dense
module, compare Eq.~(\ref{Q7}) below.  The Fourier analysis opens the
way to the analysis of scattering from quasiperiodic structures in the
same spirit as in ordinary crystallography.

\subsection{Point symmetry}\index{point symmetry}

The non-crystallographic point symmetry found in quasicrystals
produces irrational subspaces and quasiperiodicity of the Bohr class.

The discrete crystallographic point group $G$ of a lattice $\Lambda$
consists of orthogonal transformations $D(G)<O(n)$ which carry
$\Lambda$ into itself. For given dimension $n$, all the space groups
generated by pairs $(\Lambda, D(G))$ are classified by the
crystallography of $E^n$ \cite{BR}.

A point group $G$ acting by $D(G)$ on $E^m$ but incompatible with any
lattice $\Lambda \in E^m$ is called non-crystallographic.  Examples
are the cyclic group $G=C_5$ of order $5$ in $E^2$ and the icosahedral
group $G=H_3$ in $E^3$.

Mathematical schemes for aperiodic long-range order like the fivefold
symmetric Penrose rhombus tiling \cite{PE} or the icosahedral
rhombohedral tiling \cite{KN} displayed forbidden symmetries and
preceded the experiments.  The tacit assumption in physics until the
year $1984$ was that, since forbidden point symmetry was not
compatible with any periodic lattice in $E^3$, there could be no
long-range order with this point symmetry.  The discovery of
quasicrystals \cite{SBGC84} in $1984$ with icosahedral point symmetry
disproved this assumption.

After 1984, non-crystallographic point symmetry turned out to play a
key role for the understanding of new types of long-range order.  The
selection of a non-crystallographic point group generates the
irrational subspace underlying the Bohr class of quasiperiodicity.  To
explain this we first claim that for any given group $G$ of finite
order $|G|<\infty$ there exists a dimension $n$ and a lattice
embedding $\Lambda \in E^n$ with point group a representation $D(G)<
O(n)$.  To show this we recall that, in the regular $|G| \times |G|$
representation $D^{{\rm reg}}(G)$ of $G$, any element $g\in G$ is
represented by a permutation matrix.  It then follows that the regular
representation of $G$ transforms the hypercubic lattice $\mathbb{Z}^n$ into
itself and so acts as a point group of this lattice. Of course it is
technically desirable to look for the minimal lattice embedding of
$G$. This can be achieved by the use of induced representations. In
the next subsections we shall display minimal lattice embeddings.

Assume now that the $n \times n$ representation $D(G)$ admits 
a block diagonal reduction 
\begin{equation}
\label{Q6}
D(G)\sim  D'(G) \oplus D''(G),
\end{equation}
where $D'(G)$ acts non-crystallographically on $E^m=E_{\parallel}$.
Then the decomposition, Eq.~(\ref{Q4}), of $E^n$ meets the requirements
for the Bohr class of quasiperiodic functions $f^{qp}$, and moreover
these functions on $E^m$ display the non-crystallographic point
symmetry $D'(G)$.

\begin{example}[Fivefold point symmetry in $E^2$ from the root lattice
$A_4 < E^4$]
The root lattice \cite{BA}\index{root lattice} $A_4< E^4$ has as point group a 4D
representation $D(S_5)$ of $S_5$, the symmetric group of order
$5$. The cyclic subgroup $C_5<S_5$ has two inequivalent real
orthogonal 2D representations which for convenience we denote as
$D_{\parallel}(C_5)$, $D_{\perp}(C_5)$. In these two representations,
$C_5$ is generated by a rotation of angle $2\pi/5$ and $4\pi/5$,
respectively.

When the point group $D(S_5)$ of $A_4$ is restricted from $S_5$ to
$C_5$, one finds the block diagonal reduction
\begin{equation}
\label{Q6a}
D(C_5) \sim D_{\parallel}(C_5) \oplus D_{\perp}(C_5).
\end{equation}
But both 2D representations of $C_5$ are non-crystallographic,
and so the decomposition of Eq.~(\ref{Q6a}) determines two 
orthogonal irrational planes in $E^4$. The plane 
$E_{\parallel}$ is used for the Penrose and triangle quasiperiodic 
tilings discussed below.
\end{example}

On $E^m=E_{\parallel}$,
the parallel projections of the lattice points of Eq.~(\ref{Q1}),
\begin{equation}
\label{Q7}
t_{\parallel}=\sum_j m_j(a_j)_{\parallel},\quad
(m_1,m_2,\ldots,m_n)\in \mathbb{Z}^n , 
\end{equation}
form a $\mathbb{Z}$-module with basis the parallel projections of the lattice 
basis from $E^n$ to $E^m<E^n$. In the Fourier analysis, it is shown
that the Fourier amplitudes of a quasiperiodic function can be 
assigned to the reciprocal $\mathbb{Z}$-module spanned by the perpendicular
projections to $E_{\perp}$ of the reciprocal basis of $\Lambda^R$.

\subsection{Dual Voronoi and Delone cell complexes and their projection}

Dual Voronoi and Delone cell complexes in $E^n$ provide periodic
tilings of $E^n$. The T\"{u}bingen group of Kramer and coworkers since
1984 elaborated the significance of the dual cell complexes ${\cal V},
{\cal D}$ for quasiperiodicity. The boundaries of the two complexes
can be adapted to the irrational subspace $E_{\parallel}$ and its
orthogonal complement $E_{\perp}$. Parallel projections of boundaries
provide tiles, perpendicular projections their coding windows.

The cell structure of crystals in $E^n$ can, due to their periodicity, be
encoded into cells modulo lattice translations. Clearly the values of
a periodic function characterised by Eq.~(\ref{Q3a}) can be fixed on a
fundamental domain. In the language of group action, the fundamental
domain is a set of points $x \in E^n$ such that any other point of
$E^n$ can be reached from this set by a lattice vector from $\Lambda$.
This fundamental domain may be taken as the unit cell, the
parallelepiped spanned by the lattice vectors, or as the Voronoi
domain $V$, defined below in Eq.~(\ref{Q8}).

The reasoning so far puts the structure analysis of quasicrystals into
the frame of irrational embedding into a lattice and a space $\Lambda
\in E^n$.  Quasicrystals appear as crystallographic objects in
$E^n$. Visualised in $E^n$, they represent quasiperiodic sections of
dimension $m$ through periodic objects in $E^n$. By definition of
irrationality, the lattice vectors do not connect any two points on
the subspace $E^m$. It follows that a cell structure for quasicrystals
cannot be constructed modulo lattice translations. We now derive a
canonical tiling structure by projection from dual boundaries of
Voronoi and Delone cells of $\Lambda \in E^n$.  The canonical
projections from Voronoi and Delone complexes were initiated in
\cite{KN} and worked out for the fivefold symmetry in \cite{BA} and for
icosahedral point symmetry in \cite{KPZ}.  The general approach is
given in \cite{KS,KR}.

\begin{figure}\centering
\includegraphics[width=0.42\textwidth]{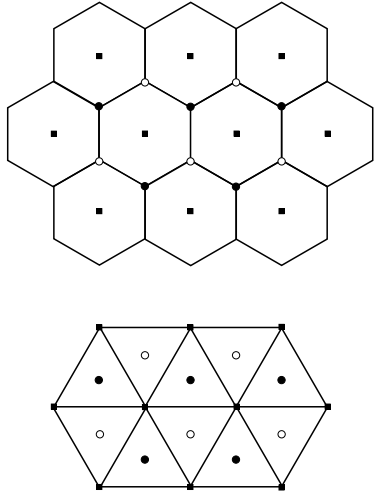}
\caption{Voronoi and dual Delone complex for the root lattice $A_2$. 
The Voronoi cells are hexagons centred at lattice points (black squares), 
the Delone cells triangles centred at two inequivalent types of
holes (black and white cycles).\label{PK:fig1}}
\end{figure}

The Voronoi domain\index{Voronoi domain} is an $n$-polytope around a
lattice point $t \in \Lambda$ defined as the set of points
\begin{equation}
\label{Q8}
V_t:= \{ x\in E^n: |x-t|\leq 
|x-t'|\; {\rm if}\; t'\neq t,\; t' \in \Lambda\}.
\end{equation}
Any Voronoi domain has a hierarchy of boundaries $X_p$ of dimension 
$p,\;0\leq p\leq n$, called $p$-boundaries. The Voronoi complex ${\cal V}$ 
is the cell complex
in $E^n$ formed by the Voronoi domains of all lattice points. It fills 
$E^n$.
Its $p$-boundaries form a hierarchy of subcomplexes of dimension 
$p,\;0\leq p\leq n$. A single fixed $p$-boundary $X_p$ determines the set
of all lattice vectors $S_{X_p}$ whose Voronoi domains contain $X_p$ 
as a boundary,
\begin{equation}
\label{Q9}
S_{X_p}:= \{ t \mid X_p \in V_t\} .
\end{equation}
We define the $(n-p)$-polytope 
dual to $X_p$ as the convex hull 
\begin{equation}
\label{Q10}
X^*_{(n-p)}:= {\rm conv}(t \in S_{X_p}).  
\end{equation}
It can be shown that $X_p$, $X^*_{(n-p)}$ have indeed complementary
dimension.  The $n$-polytopes $X^*_n$ are dual to the vertices $X_0$
of the Voronoi domains and are called the Delone cells\index{Delone
cell} of the lattice $\Lambda$.  They are centred at the vertices of
the Voronoi domains, called the holes of the lattice. These vertices
can be inequivalent under $\Lambda$ and then give rise to inequivalent
types of Delone cells.  The Voronoi cells again fill $E^n$ and form
the dual Delone cell complex ${\cal D}$ of $\Lambda$.  The dual
boundaries $X^*_{(n-p)}$, $0\leq p\leq n$, form the boundaries of the
Delone cells. Both cell complexes for the root lattice $A_2 \in E^2$
are illustrated in Fig.~\ref{PK:fig1}.\index{duality}

Given the dual boundary set from the Voronoi and Delone cell complex,
we turn to the orthogonal decomposition of $E^n$ given by
Eq.~(\ref{Q4}).  We introduce the projections of these boundaries to the
parallel and perpendicular subspaces, denoted by the subindices
$_{\parallel}, _{\perp}$. From pairs of parallel projections
${X_m}_{\parallel}$ and the orthogonal projections of their duals
${X^*_{(n-m)}}_{\perp}$ we form the set of the direct product
polytopes
\begin{equation}
\label{Q11}
T= {X_m}_{\parallel}\times {X^*_{(n-m)}}_{\perp}.
\end{equation} 
These direct product polytopes turn out to be $n$-polytopes which
provide a periodic tiling of $E^n$.\index{tiling}  By definition this periodic
tiling has the particular property that all its tiles have their
boundaries parallel or perpendicular to the subspace $E^m$.

Consider now a subspace $E^m$ with fixed value of $c_{\perp}$ according 
to Eq.~(\ref{Q5}). Its intersections with the tiling of $E^n$ 
by the direct product polytopes $T$, Eq.~(\ref{Q11}),
consists of shapes ${X_m}_{\parallel}$, which form a finite variety
of projected $m$-boundaries from the Voronoi complex ${\cal V}$
or rather from its $m$-subcomplex ${\cal V}^m$.
We call them the tiles of the quasiperiodic canonical tiling 
$({\cal T}, \Lambda)$ of $E^m$. The perpendicular projections in 
Eq.~(\ref{Q11}) are named the windows\index{window}
of the tiles. Windows can also be defined for vertices, for holes,
and for covering clusters discussed below. 
  
\begin{figure}\centering
\includegraphics[width=0.85\textwidth]{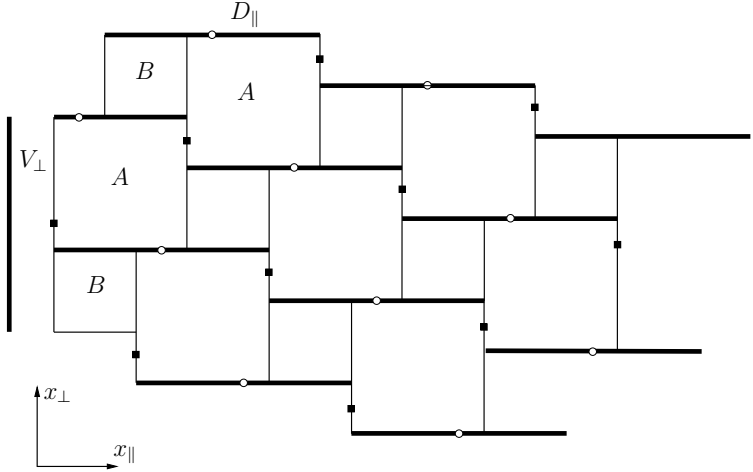}
\caption{Construction of the Fibonacci tiling.  Two new squares $A,B$
for $\Lambda=\mathbb{Z}^2$ (lattice points black squares, holes white circles)
play the role of the polytopes $T^*$ in Eq.~(\ref{Q12}). The squares are
constructed by projection from dual pairs of $1$-boundaries and tile
$E^2$ periodically. The irrational horizontal Fibonacci line
$E^1_{\parallel}$ runs through this periodic tiling. Its intersections
display the tiles $A_{\parallel}, B_{\parallel}$ of the quasiperiodic
Fibonacci tiling $({\cal T}^*, \mathbb{Z}^2)$.\label{PK:fig2}}
\end{figure}

The dual Voronoi and Delone cell complexes allow for a second construction 
with the same subspaces but an alternative to Eq.~(\ref{Q11}). 
We interchange the role of the Voronoi and Delone complex, 
consider projected pairs of boundaries 
${X^*_m}_{\parallel}$,  ${X_{(n-m)}}_{\perp}$, and form the direct 
product polytopes
\begin{equation}
\label{Q12}
T^*= {X^*_m}_{\parallel}\times {X_{(n-m)}}_{\perp}.
\end{equation} 
The intersection with $E^m$ yields as tiles the projected
$m$-boundaries of the Delone complex ${\cal D}$. This type of tiling
we denote by $({\cal T}^*,\Lambda)$. It is illustrated by the
Fibonacci tiling in Fig.~\ref{PK:fig2}.\index{Fibonacci tiling} 

More examples for these quasiperiodic tilings will be given in the next 
subsections.

\subsection{Quasiperiodic functions compatible with a tiling}

Once we have constructed a quasiperiodic tiling on the irrational
subspace $E^m < E^n$, we return to a quasiperiodic functions $f^{qp}$
of the Bohr class.  By Eq.~(\ref{Q5}) these were given as the
restrictions of $\Lambda$-periodic functions $f^p$ to their values on
the subspace $E^m$. This subspace, fixed by a perpendicular coordinate
$c_{\perp}$, slices the periodic tiles $T$ of Eq.~(\ref{Q11}) in varying
vertical positions. Starting from a general $\Lambda$-periodic
function $f^p$ we cannot infer that $f^{qp}$ obtained from it repeats
its values on separate slices of a tile of fixed shape
$X_{m{\perp}}$. In other words the functional values of a general
quasiperiodic $f^{qp}$ are still incompatible with the quasiperiodic
tiling. Compatibility is achieved by the following
construction:\index{compatibility}

Consider the periodic tiling of $E^n$ by the polytopes $T$ of
Eq.~(\ref{Q11}). Restrict a $\Lambda$-periodic function
$f^p(x)=f^p(x_{\parallel}+x_{\perp})$ on any single polytope $T$ to
functional values independent of the coordinate $x_{\perp}$.
Construct from this restricted $f^p$ by Eq.~(\ref{Q5}) the
corresponding quasiperiodic function $f^{qp}$.  It follows that the
functional values of $f^{qp}$ are repeated on tiles of the same shape
$X_{m{\parallel}}$. The quasiperiodic function $f^{qp}(x_{\parallel})$
is compatible with the tiling $({\cal T}, \Lambda)$.

From compatible functions one can now construct the notion of a
fundamental domain for quasiperiodic tilings \cite[pp.~99--100]{KP}.  As
a consequence, the functional values of a compatible quasiperiodic
function can now, as in crystals, be specified with reference to a
bounded set of points on $E^m$ modulo the projections of lattice
translations.  These concepts are put in action in the atomic
structure and the physics of quasicrystals.

\subsection{Quasiperiodic tilings from the lattices $A_4$
and $D_6$}\index{root lattice}

\begin{figure}\centering
\includegraphics[width=0.75\textwidth]{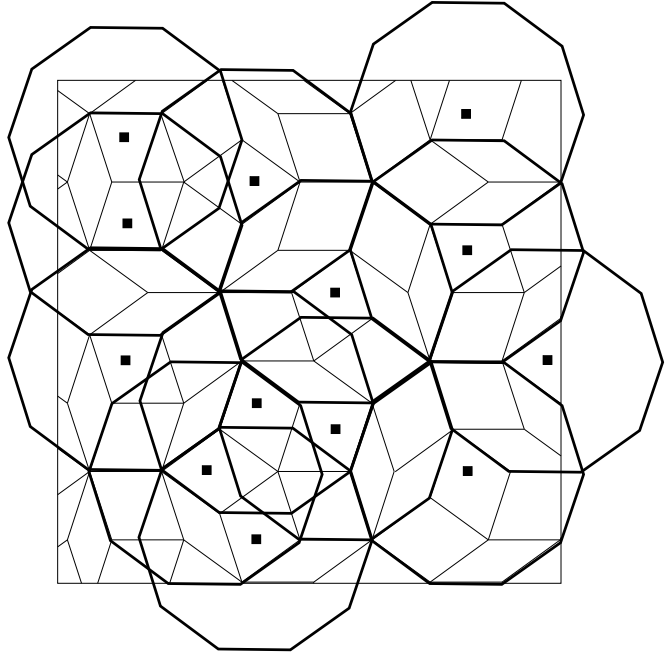}
\caption{The Penrose rhombus tiling $({\cal T}, A_4)$. It consists of
two types of rhombus tiles (thin lines), projections of $2$-boundaries
from the Voronoi complex.  Any rhombus vertex is the projection of a
hole from $\Lambda=A_4$.
In addition we show a covering of the Penrose tiling by overlapping
decagons (heavy lines) \cite{G96}. Each decagon covers $10$ rhombus tiles and is
centred at the projection of a lattice point (black square) of $A_4$.
\label{PK:fig3}\index{Penrose tiling}}
\end{figure}

We now illustrate the general constructions by examples.  The root
lattice $A_4$ admits as its point group a 4D orthogonal representation
$D(S_5)$ of the symmetric group $S_5$.  If we consider the cyclic
subgroup $C_5\in S_5$, the representation decomposes as given in
Eq.~(\ref{Q6a}) into two 2D representations.  Each of them does not
admit any periodic lattice in $E^2$.  When we construct the polytopes
according to Eq.~(\ref{Q11}), we get a quasiperiodic tiling whose
cells are projections of $2$-boundaries of the Voronoi cells of
$A_4$. These have two rhombus shapes and yield the tiling $({\cal T},
A_4)$ identical to the Penrose rhombus tiling shown in
Fig.~\ref{PK:fig3}.  Penrose \cite{PE} in 1974 constructed his rhombus
tiling without use of any projection. De Bruijn \cite{BU} proved that
it could be embedded into a 5D lattice.  Our exposition follows the
canonical minimal construction from the root lattice $A_4 \in E^4$
given in \cite{BA}.

\begin{figure}\centering
\includegraphics[width=0.7\textwidth]{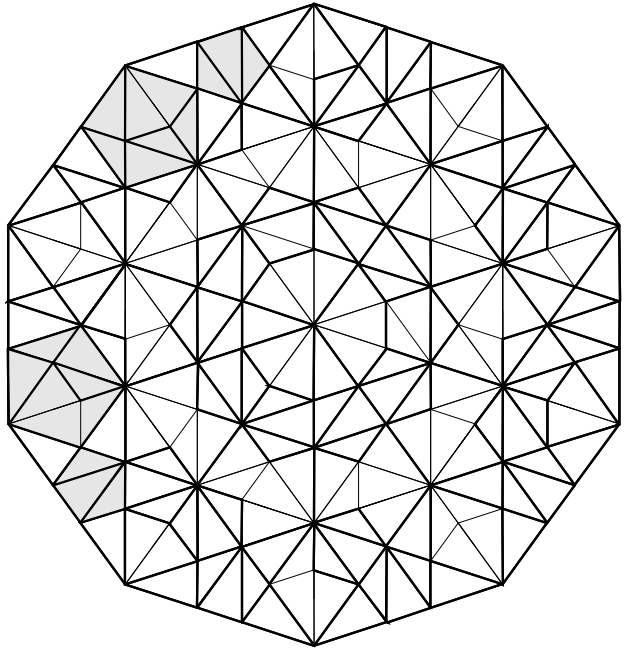}
\caption{The T\"{u}bingen triangle tiling $({\cal T}, A_4)$.  The tiles
(thin lines) are two triangular projections of Delone $2$-boundaries
from the Delone complex.  Any triangle vertex is the projection of a
lattice point.
In addition we show a covering of the triangular tiling by two types
of overlapping pentagons (heavy lines).\label{PK:fig4}\index{T\"{u}bingen 
triangle tiling}}
\end{figure}

If we use the alternative construction of Eq.~(\ref{Q12}), the tilings
of the new tiling $({\cal T}^*, A_4)$ are projections of
$2$-boundaries of the Delone complex of $A_4$. These have two triangle
shapes. The tiling is the T\"ubingen triangle tiling shown in
Fig.~\ref{PK:fig4}.

\begin{figure}\centering
\includegraphics[width=0.8\textwidth]{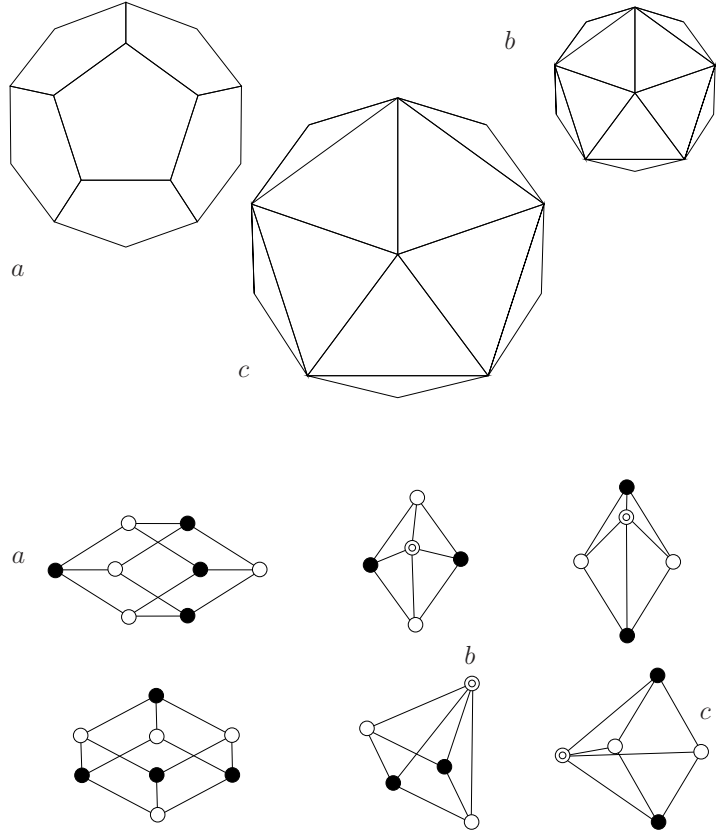}
\caption{
The 3D icosahedral tiling $({\cal T}^*, D_6)$.  Top: The three
Delone windows $D^a_{\perp},D^c_{\perp},D^b_{\perp} \in E_{\perp}$ of
the tiling $({\cal T}^*, D_6)$.  Bottom: The six tiles of this tiling
are four pyramids on a rhombus base and the two rhombohedra known from
the primitive tiling.  The holes $a,c$ are marked by black and white
circles, the holes $b$ by a double 
circle.\label{PK:fig5}\index{icosahedral tiling}}
\end{figure}

In work published before the experimental finding, Kramer and Neri
\cite{KN}\index{icosahedral tiling} in 1984 induced a 6D
representation of the icosahedral group and embedded it minimally into
the hypercubic lattice $\mathbb{Z}^6$.\index{hypercubic lattice} The
corresponding quasiperiodic tiling $({\cal T}, \mathbb{Z}^6)$ with
icosahedral point symmetry has two rhombohedral tiles, compare
Fig.~\ref{PK:fig5}.  The root lattice $D_6\in E^6$
\cite{KPZ}\index{root lattice} admits the two dual constructions
according to Eqs.~(\ref{Q11}) and (\ref{Q12}), both with icosahedral
point symmetry. These cases were worked out in detail and have been
successfully applied to icosahedral quasicrystals.\index{icosahedral
quasicrystal}\index{icosahedral tiling}

In Fig.~\ref{PK:fig5} we show the windows, compare Eq.~(\ref{Q11}) and
what follows, and the tiles of the 3D Delone-based tiling $({\cal
T}^*, D_6)$. The lattice has three types of holes denoted by
$a,b,c$. There are three corresponding types of Delone cells $D^a,
D^b, D^c$ whose perpendicular projections are shown in
Fig.~\ref{PK:fig5}.  The Delone $3$-boundaries, and their projections
which form the tiles, display as vertices these three types of holes.

\subsection{Covering of quasiperiodic tilings}\index{covering}

A new approach to quasiperiodic structure came with the idea of
covering \cite{G96}. In a covering of a tiling, every tile becomes
part of a small number of covering clusters. In contrast to a tiling,
these clusters are allowed to overlap.  Coverings of quasiperiodic
structures are treated in detail in \cite{KP}. Here we give only two
examples: The Penrose tiling $({\cal T}, A_4)$ is covered \cite{KR2}
by overlapping decagons, each consisting of $10$ rhombus tiles.  This
covering is shown in Fig.~\ref{PK:fig3}.  The triangle tiling $({\cal
T}^*, A_4)$ is covered \cite{KR2} by two types of pentagons as shown
in Fig.~\ref{PK:fig4}.

\begin{acknowledgement}
UG acknowledges support by EPSRC via Grant EP/D058465. The authors
thank D.~Shechtman and P.C.~Canfield, and the American Physical
Society, for granting permission to reproduce Figs.~\ref{UG:fig02} and 
\ref{UG:fig03} in this article.
\end{acknowledgement}

\printindex
\end{document}